\documentclass[preprint,aps,prb,groupedaddress,amssymb,showpacs]{revtex4}

\usepackage{graphicx}
\usepackage{dcolumn}
\usepackage{amsmath}
\begin{document}
\title{ High volumetric capacitance near insulator-metal  percolation transition}
\author{A. L. Efros}
\email{efros@physics.utah.edu} \affiliation{Department of Physics
\& Astronomy, University of Utah, Salt Lake City UT, 84112 USA}


\begin{abstract}
 A new type of  a capacitor   with a very high volumetric
capacitance  is proposed. It is based upon the known phenomenon of a
sharp increase of the dielectric constant of the metal-insulator
composite in the vicinity of  the percolation threshold,  but still
on the insulator side. The optimization suggests that the metallic
particles should be of nanoscale and that the distance between
planar electrodes should be somewhat larger than the correlation
length of the percolation theory and $\approx$ 10 to 20 times larger
than the size of the particles while the area of the electrodes
might be unlimited. The  random electric field in the capacitors is
found to be larger than the average field corresponding to the
potential difference of electrodes. This random field is potentially
responsible for dielectric breakdown. The estimated breakdown
voltage of the new capacitor shows that the stored energy density
might be significantly larger than that of electrolytic capacitors
while the volumetric capacitances might be comparable. The charging
and discharging times should be significantly smaller than
corresponding times of batteries and even electrolytic capacitors.
\end{abstract}
 \maketitle
  Creation of a capacitor with a very large
 capacitance per volume or weight (volumetric or  weight capacitance) is an extremely important step
 on the way to a ``green energetics". Currently the leading role in this
 field belongs to  the double-layer electrolytic capacitors also known as super- or ultra- capacitors. The
 main idea of the electrolytic capacitor is based upon a very narrow spatial
 gap between metallic electrode and electrolyte that plays the role of another
 electrode. The modern
 development, both experimental and theoretical can be found in
 the  papers\cite{pat,lar,skin,loth} and many
 others. The volumetric capacitance of these devices\cite{lar} might be about
85${\rm F/cm^3}$

 In this communication I propose a completely different
approach to the problem of a supercapacitor that according to my
estimates may successfully compete with the previous ones. The basis
of the new approach is the sharp increase of the dielectric constant
near the percolation threshold in a system of metallic nanoparticles
distributed randomly in a dielectric matrix.

Suppose that  the permittivity of the dielectric
 is $\epsilon_0 \kappa_D$ , where $\epsilon_0$ is the
permittivity of vacuum, and $\kappa_D$ is the dielectric constant.
Suppose that $x$ is a volume fraction of metal in the mixture. The
insulator-metal transition of the percolation nature occurs  at
some critical value $x_c$.  This critical fraction is practically
independent of the form of the particles and $x_c\approx 0.15$ in
three dimensional systems with a random distribution of metallic
particles\cite{book,book1}.

 It has been shown by Dubrov, Levinshtein
and Shur \cite{shur} and by Efros and Shklovskii\cite{ef}  that on
the dielectric side of the transition the dielectric constant
$\kappa$ of the  composite material with the infinite volume
diverges near the percolation threshold as
\begin{equation}\label{eps}
\kappa=\frac{\kappa_D}{\tau^q},
\end{equation}
 $\tau=x_c-x>0$, and $q$ is a critical exponent. Note that
Eq.(\ref{eps}) is only valid at $x_c-x\ll x_c$. A similar result
was obtained later by Bergman and Imry\cite{joe}.

The  origin   of the exponent $q$ is as follows. Consider a random
composite of a dielectric with a small conductivity $\sigma_D$ and
 metallic particles with a high conductivity $\sigma_M$ such that $\sigma_M\gg
\sigma_D$. One can introduce two different laws for the resulting
conductivity:
\begin{equation}\label{q}
\sigma(x)=\sigma_D/(x_c-x)^q
\end{equation}
for $x<x_c$, and
\begin{equation}\label{t}
\sigma(x)=\sigma_M(x-x_c)^t
\end{equation}
for $x>x_c$. Eq. (\ref{q}) is valid if $\sigma_D\ll\sigma(x)\ll
\sigma(x_c)$, while Eq. (\ref{t}) is valid if
$\sigma_M\gg\sigma(x)\gg \sigma(x_c)$. A smooth transition from
Eq.(\ref{q}) to Eq.(\ref{t}) occurs in a narrow region around
$x_c$ and\cite{ef}.
\begin{equation}\label{s}
\sigma(x_c)=\sigma_M(\frac{\sigma_D}{\sigma_M})^s.
\end{equation}
It has been shown\cite{ef} that the three exponents $q,t,s$ are
connected by one relation
\begin{equation}\label{rel}
q=t\left(\frac{1}{s}-1\right).
\end{equation}

Eq.(\ref{eps})  for static dielectric constant has been obtained
by considering the same problem at a finite frequency $\omega$,
doing analytical continuation\cite{ef} in the upper half plane of
a complex $\omega$, and then putting $\omega=0$. The exponent $q$
in Eq.(\ref{eps}) is the same as in Eqs.(\ref{q},\ref{rel})

  There are some controversies concerning the numerical value of
the exponent $q$  in three dimensional (3D) case. The experimental
result by Grannan {\it et al.} \cite{gran} is $q=0.73$. The same
number is given in the review paper by Clerc {\it et
al.}\cite{clerc} that summarizes the results of different
theoretical approaches. However, a similar review by Bergman and
Stroud\cite{berg} gives $q=0.76$. The last two results are close
and in this paper I make a choice in favor of experimental data
and the review \cite{clerc}, taking $q=0.73$.

The sharp increase of the dielectric constant has been confirmed
experimentally in a few other papers\cite{moya,valant}. In the
experiment\cite{moya} the dielectric constant increases about
three order of magnitude reaching $10^5$ near the percolation
threshold of molybdenum particles in a ceramic. A similar result
was reported by Pecharomman {\it et al.} \cite{pec} in ${\rm
BaTiO_{3}-Ni}$ composite. In the percolative regime they got a
high and frequency independent dielectric constant $\kappa \approx
80 000$.

The physical reason of such increase is   the existence  of
metallic clusters that have large but finite size. This size tends
to infinity near the threshold. Almost all metallic sites of such
a cluster belong to the ``dead ends" which are connected to the
rest of the cluster by one site only. The huge capacitance is
created by the small distances between sites that belong to
different metallic clusters, separated by polarized dielectric.

The percolation transition occurs when these clusters become
connected and form the infinite cluster. The characteristic size
of the finite clusters below the transition is of the order of the
correlation length
\begin{equation}\label{lc}
 L_c=l/\tau^\nu,
\end{equation}
 where $l$ is the size of a metallic particle and
$\nu$ is the exponent of the correlation length. In the 3D case
 $\nu \approx 0.87$ \cite{Stau,Sahimi}.

 Similar to Eq.(\ref{eps}),
Eq.(\ref{lc}) is
 is valid only at $x_c-x\ll x_c$. Note that both these equations and all equations obtained from them
  below are rather
 estimates because they should contain  unknown numerical
 coefficients.

In a system of finite linear size  the apparent divergence of the
dielectric constant saturates at such small values of $\tau$, when
the correlation length $L_c$ becomes of the order of  this size.
This result is common for all second-order phase transitions.

Consider a parallel plate capacitor with a large area $S$ of the
metallic plates and with a distance $L$ between the plates. The
capacitance $C=\epsilon_0 \kappa S/L$. Here $\kappa$ itself
increases with  $L$ at $L<L_c$. One can say that the dielectric
constant near the percolation threshold has a spatial dispersion
at $L<L_c$. At $L>L_c$ it is given by Eq. (\ref{eps}) and it is
independent of $L$.  Then
  the capacitance decreases with increase of $L$ at $L\gg L_c$ as $1/L$.
Thus, the optimal distance between the plates of a capacitor is
$L\approx L_c$.

However, one should keep in mind that the percolation threshold
$x_c$ has a meaning for the infinite array only.  Any set of
samples with the same size $L$ has a small critical region of
$|x_c-x|$, where the threshold may appear with the probability
about 1/2. The width of this region is\cite{book,book1} of the
order of $\delta=(l/L)^{1/\nu}$. To be sure that there are no
electrical shorts between metallic electrodes one should work on
the dielectric side outside the critical region at $\tau>\delta$.
This means that $L>L_c$. Since the numerical factor in
Eq.(\ref{lc}) is  unknown and it may depend on the shape of
metallic particles, I assume for the estimates that distance
between electrodes is $L_c$ as given by Eq.(\ref{lc}).
 To prevent  electrical shorts every
metallic electrode may be separated from infiltrated dielectric by a
thin layer of a  pure dielectric with a high electric strength.

In this approximation the capacitance per area $C_S$ has the form
\begin{equation}\label{cs}
C_S=\epsilon_0 \kappa/L_c.
\end{equation}
 It is  convenient to use  a dimensionless
length $P$ such that $L_c=Pl$, where $P=\tau^{-\nu}$. Using
Eqs.(\ref{eps},\ref{lc}) one gets
\begin{equation}\label{cs1}
C_S=\epsilon_0 \kappa_D\left(P\right)^{(q/\nu-1)}/l.
\end{equation}
 Following Clerc {\it et al.}\cite{clerc} we assume that in a 3D array the ratio
 $q/\nu=0.84$. Then
\begin{equation}\label{cs2}
C_S=\epsilon_0 \kappa_D/\left(P\right)^{0.16}l.
\end{equation}

An important characteristic of a capacitor is its volumetric
capacitance. Assuming that the distance between electrodes is
$L_c$ one gets for capacitance per volume
\begin{equation}\label{reg}
C_V=\epsilon_0 \kappa/L_c^2.
\end{equation}
 Using
Eqs.(\ref{eps},\ref{lc}) one gets

\begin{equation}\label{cv}
C_V=\epsilon_0 \kappa_D\left(\frac{L_c}{l}\right)^{q/\nu}L_c^{-2}
\end{equation}
or
\begin{equation}\label{cvp}
C_V=\epsilon_0 \kappa_D\left(P\right)^{(q/\nu-2)}/l^{2}.
\end{equation}
Assuming $q/\nu=0.84$ one gets
\begin{equation}\label{cv2}
C_V=\frac{\epsilon_0 \kappa_D}{\left(P\right)^{1.16}l^2}.
\end{equation}

 The most striking result is that near the percolation threshold
capacitances per area $C_S$ and and per volume $C_V$
 {\em depend on  the distance between
metallic electrodes  much more weakly than in the case of a regular
dielectric. } Indeed, $C_S\sim 1/L^{0.16}$ instead of $C_S\sim 1/L$
in a regular dielectric while $C_V\sim 1/L^{1.16}$ instead of
$C_V\sim 1/L^2$, where $L$ is the distance  between the plates of
the capacitor. This gain in the capacitance  results from the
spatial dispersion of the dielectric constant $\kappa$, and  it
follows formally from Eq.(\ref{eps}) that has been substituted into
Eqs.(\ref{cs},\ref{reg}).

Now we see that  close proximity to the percolation threshold is
not the optimal solution because in this case $L_c$ becomes larger
and volumetric capacitance becomes smaller.

One should keep in mind, however, that the strong enhancement of the
dielectric constant near the percolation threshold is valid if
$L_c>>l$ which also means that  $P>>1$ and $\tau^{-1}>>1$.

 The computational experience
shows that  in a 3D array  it is enough to have $P\approx$ 15 to 20.
To find $\tau$ one should use $P=1/\tau^\nu$.

Now I can do estimates. At $P=20$ and $l=1$ nm one gets

\begin{equation}\label{cs3}
 C_S\approx 0.55 \frac{\kappa_D}{100}{\rm F/m^2}
\end{equation}

and

\begin{equation}\label{cvq}
 C_V=27\frac{\kappa_D}{100}{\rm F/cm^3}.
\end{equation}

\begin{figure}
  \includegraphics[width=7 cm]{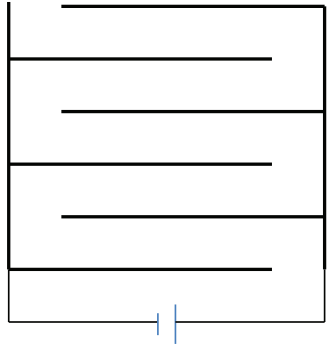}\\
  \caption{Cross-section of a   capacitor consisting of five thin capacitors connecting parallel.
  Straight thick lines
  show metallic
  electrodes. All space between them filled with dielectric infiltrated by metallic
  nanoparticles except thin preventing layer of a pure dielectric
  around each metallic electrode (see text).
  Thin lines show the battery connection to the capacitor.
  The openings between horizontal and vertical lines are shown large
  for clarity only. In fact, they should be  large enough to prevent
  electrical shorts  only.
  }
\end{figure}

 Note that
the value of $\kappa_D$ in this estimate might be significantly
larger than 100. For example  recently\cite{col} $\kappa_D=3000$
has been discovered in ${\rm Sr_{1-x}Pr_{x}TiO_3}$ ceramics.
Taking this value one finds that $C_V=810{\rm F/cm^3}$. Thus this
estimate shows that the proposed capacitor might be comparable
with the electrolytic one that provides $C_V=85{\rm F/cm^3}$.

The  capacitor under discussion  is  very thin though the area of
the metallic electrodes might be large. It is possible to make a
capacitor of a cubic or of any other form with a parallel connection
of the thin capacitors. An example of such a capacitor is shown in
Fig. 1. It has almost the same volumetric capacitance as each thin
capacitor.

 Now I discuss the important issue of the maximum energy stored
in the capacitor at a given capacitance. This energy is limited by
the value of the  voltage that can be  safely applied to the
capacitors without causing  dielectric breakdown.

 One can show that  a capacitor
 made from a dielectric filled by
 metallic particles in the vicinity of  the percolation threshold has a
  microscopic field ${\bf e}$ of  random orientation
 and  much larger than the macroscopic  field $E=U/L_c$, calculated
 from the applied voltage $U$.

   Now I estimate the value of the  field ${\bf e}$. The energy $Q$ of
the capacitor per unit surface area is
\begin{equation}\label{Qu}
Q=C_SU^2/2.
\end{equation}
  On the other hand,
this energy can be calculated microscopically as the energy of
 electric field ${\bf e}$ per unit surface area of the capacitor
\begin{equation}\label{Qs}
Q=\frac{\epsilon_0\kappa_D}{2L_c^2}\int ({\bf e})^2 dV,
\end{equation}
where the integral is taken over the cube $L_c^3$, but the
integrand is non-zero in the  dielectric regions only because the
field inside metals is zero. Note that Eq.(\ref{Qs}) contains
dielectric constant $\kappa_D$ of a pure dielectric. Increase of
the macroscopic dielectric constant $\kappa$ and of the
capacitance is due to the microscopic field ${\bf e}$. This field
is the solution of the Poisson equation that takes into account
metallic infiltration. Introducing the squared field averaged over
the cube $L_c^3$
\begin{equation}\label{sq}
 <({\bf e})^2>=(1/L_c^3)\int ({\bf e})^2 dV,
\end{equation}
one can write the  energy of the capacitor $Q$ per unit area in
the form
\begin{equation}\label{Q}
Q=\frac{\epsilon_0\kappa_D}{2} <({\bf e})^2> L_c.
\end{equation}

  Comparing Eqs.
 (\ref{Qu},\ref{cs1}), and (\ref{Q}) one  gets
\begin{equation}\label{e}
e\approx \sqrt{<({\bf
e})^2>}=\frac{U}{\sqrt{L_cl}}\left(L_c/l\right)^{(q/\nu-1)/2}.
\end{equation}
Note that
\begin{equation}\label{co}
e/E\approx (L_c/l)^{q/2\nu}>>1.
\end{equation}

 Strictly speaking   the  field $e$ is inhomogeneous, but the
inhomogeneity is not strong. The microscopic field and the large
value of the permittivity near the percolation threshold are created
by the dead ends of the  finite clusters. The large capacitance can
be understood in terms of the large areas covered by the dead ends
of different finite clusters with different potentials. This
coverage is dense so that the microscopic field should not be
strongly inhomogeneous. Therefore I think that the fluctuations of
the field $e$ are not very important, and this field can be used to
estimate  the breakdown field.

One should also take into account that an  electrical short
between two finite clusters that appears due to a fluctuation
means only that they become one cluster. Only multiple shorts may
create the percolation between the electrodes that destroys the
device.

 To estimate the breakdown
voltage $U_b$  assume that the breakdown in the dielectric occurs
due to  tunnelling current which is usually the main reason at small
distances\cite{brake}. The  field ${\bf e}$  changes its direction
at a distance of a few values of $l\approx 1{\rm nm}$. Then the
typical voltage drop inside the dielectric is of the order of
\begin{equation}\label{d}
e l=l\frac{U}{P^{(1-q/2\nu)}}.
\end{equation}

 The tunnelling current appears when the voltage drop  given by the left hand  side of
  Eq.(\ref{d}) is equal to the
 gap width of the dielectric $E_g$ (in volts). Due to the fluctuations
 of the field $e$ it might  be somewhat less than $E_g$,
 say $E_g/\sqrt{2}$. Then from Eq.(\ref{d}) one gets
 \begin{equation}\label{br}
U_b=\frac{E_g}{\sqrt{2}}P^{(1-q/2\nu)}.
\end{equation}
At  $P=20$ and   $E_g=3{\rm V}$ the maximum voltage that can be
applied is $U_b=12{\rm V}$.

These calculations reveal another important advantage of the
proposed capacitor. The breakdown voltage is larger than
$E_g/\sqrt{2}$ in volts. Note that the random field $e$ that
creates breakdown is larger than the average field $E$, but the
tunnelling distance $l$ is so small that the voltage drop $el$ is
smaller than potential difference $U$ that is applied to a
capacitor (See Eq.(\ref{d})).

Now we can estimate the maximum density of the stored energy $W$
as
\begin{equation}\label{W}
W=\frac{C_V U_b^2}{2}.
\end{equation}
Using Eqs(\ref{cvp},\ref{br}) one gets
\begin{equation}\label{wr}
W=\frac{9}{4}\frac{\epsilon_0 \kappa_D}{l^2}(E_g/3V)^2.
\end{equation}
 This equation does not contain any critical exponents. If
$E_g=3V$, $l=1{\rm nm}$, one finds $W=0.55(\kappa_D/100) {\rm
Wh/cm^3}$. This is a very large volumetric energy. For the
electrolytic capacitors the  energy per weight is about 5 Wh/kg
(See \cite{pat}). This happens because the breakdown voltage in a
double-layer electrolytic capacitor is only
 (2 to 3)V.

 Note that the charging and discharging times of the proposed
capacitor should be much smaller than the corresponding times of
 batteries and even smaller than times of electrolytic capacitors. The reason is that
 both   batteries and   electrolytic capacitors operate with the  ion currents. In
the proposed capacitor  only the time of polarization of the
dielectric is relevant.

One can think that due to the metal infiltration the leakage current
in the proposed capacitor should be larger than in a regular
dielectric based capacitor. In principle, this is true, but the
effect is not strong. The conductivity of the infiltrated system
near the percolation threshold is given by Eq.(\ref{s}). It follows
from Eq.(\ref{rel}) that $s=0.72$. Then
\begin{equation}\label{si}
\sigma(x_c)=\sigma_D^{0.72} \sigma_M^{0.28}.
\end{equation}
Thus, the influence of metallic particles is not strong. This might
be the reason why Pecharroman and Moya\cite{moya} claimed a low
leakage current in their experiment.

 In summary,  as follows from my estimates,
 dielectric media with metallic nanoparticles that are near the
 percolation threshold might be  promising materials for capacitors
 with  large volumetric capacitance and  large stored energy density.
The material parameters I use for the estimates are far from
exceptional. They
 do not correspond to a single specific
material. I think that success or failure of the whole idea depends
on the proper choice of the dielectric material. The most important
requirements are high dielectric constant and possibility of
infiltrating by metallic nanoparticles using an advanced
nanotechnology. The requirement for the gap $E_g$ depends on whether
one needs high voltage or low voltage capacitor. If volumetric
capacitance is large enough, the stored energy density might be
large in both cases.

  I am grateful to   C. Boehme, M. Reznikov,  B. I. Shklovskii,  M. S. Shur, and John Worlock for valuable discussions and advises.

\bibliography{cap}

\begin{thebibliography}{19}
\expandafter\ifx\csname natexlab\endcsname\relax\def\natexlab#1{#1}\fi
\expandafter\ifx\csname bibnamefont\endcsname\relax
  \def\bibnamefont#1{#1}\fi
\expandafter\ifx\csname bibfnamefont\endcsname\relax
  \def\bibfnamefont#1{#1}\fi
\expandafter\ifx\csname citenamefont\endcsname\relax
  \def\citenamefont#1{#1}\fi
\expandafter\ifx\csname url\endcsname\relax
  \def\url#1{\texttt{#1}}\fi
\expandafter\ifx\csname urlprefix\endcsname\relax\def\urlprefix{URL }\fi
\providecommand{\bibinfo}[2]{#2}
\providecommand{\eprint}[2][]{\url{#2}}

\bibitem[{\citenamefont{Simon and Gogotsi}(2008)}]{pat}
\bibinfo{author}{\bibfnamefont{P.}~\bibnamefont{Simon}} \bibnamefont{and}
  \bibinfo{author}{\bibfnamefont{Y.}~\bibnamefont{Gogotsi}},
  \bibinfo{journal}{Nature Materials} \textbf{\bibinfo{volume}{7}},
  \bibinfo{pages}{845} (\bibinfo{year}{2008}).

\bibitem[{\citenamefont{Largeot et~al.}(2008)\citenamefont{Largeot, Portel,
  Chmiola, Taberna, Gogotsi, and Simon}}]{lar}
\bibinfo{author}{\bibfnamefont{C.}~\bibnamefont{Largeot}},
  \bibinfo{author}{\bibfnamefont{C.}~\bibnamefont{Portel}},
  \bibinfo{author}{\bibfnamefont{J.}~\bibnamefont{Chmiola}},
  \bibinfo{author}{\bibfnamefont{P.-L.} \bibnamefont{Taberna}},
  \bibinfo{author}{\bibfnamefont{Y.}~\bibnamefont{Gogotsi}}, \bibnamefont{and}
  \bibinfo{author}{\bibfnamefont{P.}~\bibnamefont{Simon}}, \bibinfo{journal}{J.
  Am. Chem. Soc.} \textbf{\bibinfo{volume}{69}}, \bibinfo{pages}{2730}
  (\bibinfo{year}{2008}).

\bibitem[{\citenamefont{Skinner et~al.}(2010)\citenamefont{Skinner, Loth, and
  Shklovskii}}]{skin}
\bibinfo{author}{\bibfnamefont{B.}~\bibnamefont{Skinner}},
  \bibinfo{author}{\bibfnamefont{M.~S.} \bibnamefont{Loth}}, \bibnamefont{and}
  \bibinfo{author}{\bibfnamefont{B.~I.} \bibnamefont{Shklovskii}},
  \bibinfo{journal}{Phys. Rev. Lett.} \textbf{\bibinfo{volume}{104}},
  \bibinfo{pages}{128302} (\bibinfo{year}{2010}).

\bibitem[{\citenamefont{M.S.Loth et~al.}(2010)\citenamefont{M.S.Loth, Skinner,
  and Shklovskii}}]{loth}
\bibinfo{author}{\bibnamefont{M.S.Loth}},
  \bibinfo{author}{\bibfnamefont{B.}~\bibnamefont{Skinner}}, \bibnamefont{and}
  \bibinfo{author}{\bibfnamefont{B.~I.} \bibnamefont{Shklovskii}},
  \bibinfo{journal}{Phys. Rev. E} \textbf{\bibinfo{volume}{82}},
  \bibinfo{pages}{056102} (\bibinfo{year}{2010}).

\bibitem[{\citenamefont{Shklovskii and Efros}(1984)}]{book}
\bibinfo{author}{\bibfnamefont{B.~I.} \bibnamefont{Shklovskii}}
  \bibnamefont{and} \bibinfo{author}{\bibfnamefont{A.~L.} \bibnamefont{Efros}},
  \emph{\bibinfo{title}{Electronic Properties of Doped Semiconductors}}
  (\bibinfo{publisher}{Springer-Verlag}, \bibinfo{year}{1984}),
  \bibinfo{note}{chapter 5}.

\bibitem[{\citenamefont{Efros}(1986)}]{book1}
\bibinfo{author}{\bibfnamefont{A.~L.} \bibnamefont{Efros}},
  \emph{\bibinfo{title}{Physics and Geometry of Disorder}}
  (\bibinfo{publisher}{Mir, Moscow}, \bibinfo{year}{1986}),
  \bibinfo{note}{chapter 9}.

\bibitem[{\citenamefont{Dubrov et~al.}(1976)\citenamefont{Dubrov, Levinshtein,
  and Shur}}]{shur}
\bibinfo{author}{\bibfnamefont{V.~E.} \bibnamefont{Dubrov}},
  \bibinfo{author}{\bibfnamefont{M.~E.} \bibnamefont{Levinshtein}},
  \bibnamefont{and} \bibinfo{author}{\bibfnamefont{M.~S.} \bibnamefont{Shur}},
  \bibinfo{journal}{Sov. Phys.JETP} \textbf{\bibinfo{volume}{43}},
  \bibinfo{pages}{1050} (\bibinfo{year}{1976}).

\bibitem[{\citenamefont{Efros and Shklovskii}(1976)}]{ef}
\bibinfo{author}{\bibfnamefont{A.~L.} \bibnamefont{Efros}} \bibnamefont{and}
  \bibinfo{author}{\bibfnamefont{B.~I.} \bibnamefont{Shklovskii}},
  \bibinfo{journal}{Phys.Status. Solidi} \textbf{\bibinfo{volume}{76}},
  \bibinfo{pages}{475} (\bibinfo{year}{1976}).

\bibitem[{\citenamefont{Bergman and Imry}(1977)}]{joe}
\bibinfo{author}{\bibfnamefont{D.}~\bibnamefont{Bergman}} \bibnamefont{and}
  \bibinfo{author}{\bibfnamefont{Y.}~\bibnamefont{Imry}},
  \bibinfo{journal}{Phys. Rev. Lett} \textbf{\bibinfo{volume}{39}},
  \bibinfo{pages}{1222} (\bibinfo{year}{1977}).

\bibitem[{\citenamefont{Grannan et~al.}(1981)\citenamefont{Grannan, Garland,
  and Tanner}}]{gran}
\bibinfo{author}{\bibfnamefont{D.~M.} \bibnamefont{Grannan}},
  \bibinfo{author}{\bibfnamefont{J.~C.} \bibnamefont{Garland}},
  \bibnamefont{and} \bibinfo{author}{\bibfnamefont{D.~B.}
  \bibnamefont{Tanner}}, \bibinfo{journal}{Phys. Rev. Lett}
  \textbf{\bibinfo{volume}{46}}, \bibinfo{pages}{375} (\bibinfo{year}{1981}).

\bibitem[{\citenamefont{Clerc et~al.}(1990)\citenamefont{Clerc, Giraud,
  Laugier, and Luck}}]{clerc}
\bibinfo{author}{\bibfnamefont{J.}~\bibnamefont{Clerc}},
  \bibinfo{author}{\bibfnamefont{G.}~\bibnamefont{Giraud}},
  \bibinfo{author}{\bibfnamefont{J.~M.} \bibnamefont{Laugier}},
  \bibnamefont{and} \bibinfo{author}{\bibfnamefont{J.~M.} \bibnamefont{Luck}},
  \bibinfo{journal}{Adv.Phys.} \textbf{\bibinfo{volume}{39}},
  \bibinfo{pages}{191} (\bibinfo{year}{1990}).

\bibitem[{\citenamefont{Bergman and Stroud}(1992)}]{berg}
\bibinfo{author}{\bibfnamefont{D.~J.} \bibnamefont{Bergman}} \bibnamefont{and}
  \bibinfo{author}{\bibnamefont{Stroud}}, \emph{\bibinfo{title}{Solid State
  Physics}}, vol.~\bibinfo{volume}{46} (\bibinfo{publisher}{Academic Press
  inc.}, \bibinfo{year}{1992}).

\bibitem[{\citenamefont{Pecharroman and Moya}(2000)}]{moya}
\bibinfo{author}{\bibfnamefont{C.}~\bibnamefont{Pecharroman}} \bibnamefont{and}
  \bibinfo{author}{\bibfnamefont{J.~S.} \bibnamefont{Moya}},
  \bibinfo{journal}{Adv. Materials} \textbf{\bibinfo{volume}{12(4)}},
  \bibinfo{pages}{294} (\bibinfo{year}{2000}).

\bibitem[{\citenamefont{Valant et~al.}(2006)\citenamefont{Valant, Dakskobler,
  Ambrozic, and Kosmac}}]{valant}
\bibinfo{author}{\bibfnamefont{M.}~\bibnamefont{Valant}},
  \bibinfo{author}{\bibfnamefont{A.}~\bibnamefont{Dakskobler}},
  \bibinfo{author}{\bibfnamefont{M.}~\bibnamefont{Ambrozic}}, \bibnamefont{and}
  \bibinfo{author}{\bibfnamefont{T.}~\bibnamefont{Kosmac}},
  \bibinfo{journal}{J. European Ceramic Soc.} \textbf{\bibinfo{volume}{26}},
  \bibinfo{pages}{891} (\bibinfo{year}{2006}).

\bibitem[{\citenamefont{Pecharroman et~al.}(2001)\citenamefont{Pecharroman,
  Esteban-Betegon, Bartolome, Lopez-Esteban, and Moya}}]{pec}
\bibinfo{author}{\bibfnamefont{C.}~\bibnamefont{Pecharroman}},
  \bibinfo{author}{\bibfnamefont{F.}~\bibnamefont{Esteban-Betegon}},
  \bibinfo{author}{\bibfnamefont{J.~F.} \bibnamefont{Bartolome}},
  \bibinfo{author}{\bibfnamefont{S.}~\bibnamefont{Lopez-Esteban}},
  \bibnamefont{and} \bibinfo{author}{\bibfnamefont{J.~S.} \bibnamefont{Moya}},
  \bibinfo{journal}{Adv.Materials} \textbf{\bibinfo{volume}{13}},
  \bibinfo{pages}{1541} (\bibinfo{year}{2001}).

\bibitem[{\citenamefont{Stauffer and Aharony}(1992)}]{Stau}
\bibinfo{author}{\bibfnamefont{D.}~\bibnamefont{Stauffer}} \bibnamefont{and}
  \bibinfo{author}{\bibfnamefont{A.}~\bibnamefont{Aharony}},
  \emph{\bibinfo{title}{Introduction to percolation theory}}
  (\bibinfo{publisher}{Taylor\&Francis}, \bibinfo{year}{1992}),
  \bibinfo{note}{p. 52}.

\bibitem[{\citenamefont{Sahimi}(1990)}]{Sahimi}
\bibinfo{author}{\bibfnamefont{M.}~\bibnamefont{Sahimi}},
  \emph{\bibinfo{title}{Applications of percolation theory}}
  (\bibinfo{publisher}{Taylor\&Francis}, \bibinfo{year}{1990}),
  \bibinfo{note}{p. 16}.

\bibitem[{\citenamefont{Liu et~al.}(2010)\citenamefont{Liu, Liu, ping Zhou, He,
  na~Su, Cao, and wu~Zhang}}]{col}
\bibinfo{author}{\bibfnamefont{C.}~\bibnamefont{Liu}},
  \bibinfo{author}{\bibfnamefont{P.}~\bibnamefont{Liu}},
  \bibinfo{author}{\bibfnamefont{J.}~\bibnamefont{ping Zhou}},
  \bibinfo{author}{\bibfnamefont{Y.}~\bibnamefont{He}},
  \bibinfo{author}{\bibfnamefont{L.}~\bibnamefont{na~Su}},
  \bibinfo{author}{\bibfnamefont{L.}~\bibnamefont{Cao}}, \bibnamefont{and}
  \bibinfo{author}{\bibfnamefont{H.}~\bibnamefont{wu~Zhang}},
  \bibinfo{journal}{J. Appl. Phys.} \textbf{\bibinfo{volume}{107}},
  \bibinfo{pages}{094108} (\bibinfo{year}{2010}).

\bibitem[{\citenamefont{Lin et~al.}(2005)\citenamefont{Lin, Ye, and
  Wilk}}]{brake}
\bibinfo{author}{\bibfnamefont{H.~C.} \bibnamefont{Lin}},
  \bibinfo{author}{\bibfnamefont{P.~D.} \bibnamefont{Ye}}, \bibnamefont{and}
  \bibinfo{author}{\bibfnamefont{G.~D.} \bibnamefont{Wilk}},
  \bibinfo{journal}{Appl. Phys. Lett} \textbf{\bibinfo{volume}{87}},
  \bibinfo{pages}{182904} (\bibinfo{year}{2005}).

\end{thebibliography}
\end{document}